\documentclass[12pt,a4paper]{article}
\usepackage[utf8]{inputenc}
\usepackage[utf8]{inputenc}
\usepackage[T1]{fontenc}
\usepackage{amsmath}
\usepackage{amssymb}
\usepackage{graphicx}
\usepackage{lineno}
\usepackage{geometry}
\usepackage{dsfont}
\usepackage{mathtools}
\usepackage{tgtermes}
\numberwithin{equation}{section}
\usepackage{authblk}
\usepackage{hyperref} 
\usepackage{cite}

\newcommand{\be}{\begin{equation}}
\newcommand{\ee}{\end{equation}}
\newcommand{\ba}{\begin{array}{lll}}
\newcommand{\ear}{\end{array}}

\def\sp{\mathrm{Span}}
\def\bJ{\boldsymbol{J}}
\def\bP{\boldsymbol{P}}
\def\bZ{\boldsymbol{Z}}
\def\bt{\boldsymbol{t}}
\def\btau{\boldsymbol{\tau}}
\def\bG{\boldsymbol{G}}
\def\bH{\boldsymbol{H}}
\def\bM{\boldsymbol{M}}
\def\bS{\boldsymbol{S}}

\def\bT{\boldsymbol{T}}

\def\bA{\boldsymbol{A}}
\def\ed{\mathrm{d}}

\def\nw{\mathfrak{nw}}
\def\pnw{\tilde{\mathfrak{nw}}}

\def\bal{\boldsymbol{a}}
\def\bad{\boldsymbol{a}^{\dagger}}
\def\bI{\boldsymbol{I}}
\def\bN{\boldsymbol{N}}

\def\bX{\boldsymbol{X}}
\def\bY{\boldsymbol{Y}}

\vspace{-10pt}
\date{}

\begin{document}
\title{\textbf{Non-relativistic symmetries in three space-time dimensions and the Nappi-Witten algebra}}

\author[]{
Diego M. Pe\~{n}afiel\thanks{diego.molina.p@pucv.cl}
\,\;\&\;
Patricio Salgado-Rebolledo\thanks{patricio.salgado@pucv.cl}}

\affil[]{Instituto de F\'isica Pontificia Universidad Cat\'olica de Valpara\'iso Casilla 4059, Valpara\'iso, Chile}

\maketitle

\begin{abstract}
We show that the Extended Bargmann and Newton-Hooke algebras in 2+1 dimensions can be obtained as expansions of the Nappi-Witten algebra. The result can be generalized to obtain two infinite families of non-relativistic symmetries, which include the Maxwellian Exotic Bargmann symmetry, its generalized Newton-Hooke counterpart, and its Hietarinta dual. In each case, the invariant bilinear form on the Nappi-Witten algebra leads to the invariant tensor on the expanded algebra, allowing one to construct the corresponding Chern-Simons gravity theory. \end{abstract}

\newpage
\tableofcontents
\newpage

\section{Introduction}

Non-relativistic (NR) gravity theories have received considerable attention in the last years due to their relation with non-AdS extensions of holography \cite{Nishida:2007pj,Taylor:2008tg,Herzog:2008wg,Balasubramanian:2008dm,Volovich:2009yh,Janiszewski:2012nb,Bagchi:2012cy,Christensen:2013rfa,Dobrev:2013kha,Hartong:2014oma,Christensen:2013lma}, which play an important role in the description of strongly coupled condensed matter systems \cite{Duval:2001hu,Horvathy:2004am,Son:2005rv,Son:2008ye,Hoyos:2011ez,Son:2013rqa,Abanov:2014ula,Gromov:2014vla,Geracie:2014nka}.

In the NR limit, the gravitational interaction is described in a space-covariant way in terms of Newton-Cartan geometry \cite{cartan1923varietes,cartan1924varietes,Duval:1984cj,Duval:1993pe}. Such gravity theory is obtained by gauging the Bargmann algebra \cite{Andringa:2010it}. In 2+1 dimensions, Newton-Cartan gravity can be formulated as a Chern-Simons theory invariant under the Bargmann algebra \cite{Papageorgiou:2009zc,Bergshoeff:2016lwr} (see also \cite{Matulich:2019cdo} for a general classification of non-relativistic limits of three-dimensional Einstein gravity). This can be generalized to the case of the Newton-Hooke and Schr\"odinger symmetries, whose gravity theories lead to different generalization of Horava-Lifshitz gravity in 2+1 dimensions \cite{Papageorgiou:2010ud,Hartong:2016yrf} (for a relativistic reinterpretation of the three-dimensional Schr\"odinger symmetry and the corresponding Chern-Simons theory, see \cite{Chernyavsky:2019hyp}). Moreover, the Bargmann-invariant Chern-Simons theory can be non-centrally extended to include a covariantly constant background field \cite{Aviles:2018jzw}, leading to Maxwellian Exotic Bargmann gravity.

Lie algebra expansion methods \cite{Hatsuda:2001pp,deAzcarraga:2002xi,Izaurieta:2006zz,deAzcarraga:2007et} provide a powerful tool to analyse the relation between different symmetries and the physical systems that are invariant under them. Recently, it was shown that NR symmetries can be obtained by expanding the Poincar\'e algebra, and that gravitational theories invariant under these algebras can be constructed as expansions of the Einstein-Hilbert action \cite{Bergshoeff:2019ctr}. Along the same lines, three-dimensional Bargmann supergravity has been found by expanding a Chern-Simons action invariant under the super Poincar\'e algebra \cite{deAzcarraga:2019mdn}. In \cite{Gonzalez:2016xwo}, the expansion of Galilean symmetries was carried out, leading to NR limits of the generalized Poincar\'e and generalized AdS symmetries \cite{Concha:2014zsa,Concha:2016kdz}. Remarkably, expanded Galilean algebras in 3+1 dimensions allow one to recover the MOND model proposed by Bekenstein and Milgrom in \cite{Bekenstein:1984tv} as an alternative dark matter \cite{Rubio:2018itx}.

In this article, we show that NR symmetries in 2+1 dimensions can be obtained as expansions of the universal central extension of the two-dimensional Euclidean algebra, also known as the Nappi-Witten algebra \cite{Nappi:1993ie}. In doing this, we will recover several symmetries known in the literature such as the Extended Bargmann, Extended Newton-Hooke, Carroll, and Maxwellian Exotic Bargmann algebras together with their corresponding Chern-Simons actions. Furthermore, this procedure allows one to obtain novel symmetries such as the NR limit of the semi-simple extension of the Poincar\'e algebra in 2+1 dimensions \cite{Gomis:2009dm,Soroka:2011tc}, the NR limit of the bosonic Hietarinta algebra studied in \cite{Bansal:2018qyz,Chernyavsky:2019hyp}, and NR versions of the generalized Poincar\'e and generalized AdS symmetries introduced in \cite{Concha:2014zsa,Concha:2016kdz}. 

Moreover, we will use the fact that Nappi-Witten algebra is a double extension\footnote{Double extensions have been used to study several NR symmetries in 2+1 dimensions in a unified fashion \cite{Matulich:2019cdo}, and to construct extended stringy Galilean symmetries \cite{Aviles:2019xed}.} of the two-dimensional translation algebra \cite{Figueroa-OFarrill:1994liu} to define in an analogue way the double extension the translation algebra on Minkowski space \cite{Cangemi:1992up,Cangemi:1992bj}. We will show that the expansion method can be used on algebra to obtain generalizations of the pseudo-Newton-Hooke symmetry discussed in \cite{Hartong:2017bwq}.

The paper is organized as follows: In Section \ref{expnw}, we show how the Nappi-Witten algebra can be expanded using the method based on Abelian semigroups \cite{Izaurieta:2006zz}. In Section \ref{genbarg}, we construct generalized Bargmann algebras as a particular case of expanded Nappi-Witten algebras. In Section \ref{gennh}, an analogue procedure is followed to construct generalized Newton-Hooke symmetries, which lead to the generalized Bargmann case upon contraction. In Section \ref{carroll}, we show how to construct Carroll gravity by means of a non-standard semigroup expansion. In Section \ref{pseudo}, we obtain the pseudo-Newton-Hooke symmetry by expanding the central extension of the Poincar\'e algebra in 1+1 dimensions and show how to generalize this construction to obtain more general symmetries of this type. We conclude in Section \ref{conclusions} giving a summary of our results, discuss its implications and comment on possible future directions.

Note added: while this article was in preparation, it came to our knowledge a new manuscript by  P. Concha and E. Rodr\'{i}guez, which presents some overlap with particular cases of our results \cite{Concha:2019lhn}.


\section{Expansions of the Nappi-Witten algebra}
\label{expnw}
The Nappi-Witten algebra $\nw$ is given by the four-dimensional Heisenberg algebra
\be\label{nw1}
\left[\bal,\bad\right]=\bI\,,\quad
\left[\bN,\bad\right]=\bad\,,\quad
\left[\bN,\bal\right]=-\bal\,,
\ee
and it is isomorphic to the universal central extension of the Euclidean algebra in two-dimensions \cite{Figueroa-OFarrill:1999cmq}. In fact, by making the redefinitions
\be
\bt_1=-i(\bal+\bad)\,,\quad
\bt_2=\bal-\bad\,,\quad
\bt=-i\bN\,,\quad
\btau=2i\bI\,,
\ee
the commutation relations \eqref{nw1} are brought to the form \cite{Nappi:1993ie}
\be\label{nwalg}
\left[\bt,\bt_{a}\right]=\epsilon_{a}^{\;\;b}\,\bt_{b}\,,\quad\left[\bt_{a},\bt_{b}\right]=\epsilon_{ab}\,\btau \,,
\ee
where now $\bt$ is interpreted as the generator of rotations, $\bt_a=\left\{\bt_1,\bt_2 \right\}$ stands for the generators of translations on the plane, and $\btau$ is a central term.
As the Nappi-Witten algebra is not semi-simple, its Killing form is degenerate. However, it admits the following non-degenerate invariant bilinear form
\be\label{nwmetric}
    \langle \bt_a\, \bt_b\rangle = \delta_{ab}\,,\quad
    \langle \bt\,\btau\rangle = 1 \,,\quad
    \langle \bt\,\bt\rangle = \nu\, . 
\ee
In the following, it will be of use to consider the Chern-Simons form
\be\label{csform}
\mathcal{L}_{CS}=\langle \bA \wedge \ed \bA+\frac{2}{3}\bA\wedge\bA\wedge\bA\rangle\,,
\ee
which can be constructed out of the invariant form \eqref{nwmetric} and a $\nw$-valued connection one-form
\be
\bA=\theta^a \bt_a+\alpha \bt+\beta\btau \,.
\ee
Putting all together the Chern-Simons form can be written as
\be
\mathcal{L}_{\nw}=\theta^a \wedge R_a+\alpha\wedge \ed\beta+\beta\wedge \ed\alpha +\nu\,\alpha \wedge \ed\alpha \,,
\ee
where $R^a$ is the one-form curvature associated to the gauge field $\theta^a$, and reads
\be\label{curvature}
R^a=\ed\theta^a+\epsilon^{a}_{\;\;b}\,\theta^b\wedge\alpha\,.
\ee

\subsection{Expanding $\nw$}
\label{exph}

Let $S=\left\{ \lambda _{0},\lambda _{1},\dots
,\lambda _{N}\right\} $ be a finite Abelian semigroup endowed with an associative product of the form
\begin{equation}\label{slaw}
\lambda_i \cdot \lambda_j = \lambda_{i\diamond j}\,,
\end{equation}%
where $\diamond$ characterizes the semigroup product law. Then, following \cite{Izaurieta:2006zz}, the direct product of this semigroup with the $\nw$ algebra
\be\label{expandedalg}
S\times\nw = \sp\big\{ S\otimes  \left\{\bt,\bt_a,\btau  \right\}\big\}\,,
\ee
is also a Lie algebra and defines an expansion of $\nw$ by $S$. The generators of the expanded algebra can be explicitly written as
\be\label{expandedgen}
\bt^{(i)}_a=\lambda_i \otimes\bt_{a}\,,\quad
\bt^{(i)}=\lambda_i\otimes\bt \,,\quad
\btau^{(i)} =\lambda_i \otimes\btau\,,
\ee
and satisfy the commutation relations
\be
\left[ \bt^{(i)},\bt^{(j)}_a\right] = \epsilon_{a}^{\;\;b} \bt^{(i\diamond j)}_b \, , \quad
\left[ \bt^{(i)}_a,\bt^{(j)}_b\right] = \epsilon_{a}^{\;\;b} \btau^{(i\diamond j)} \, . \label{SexpCom}
\ee

Using \eqref{nwmetric}, an non-degenerate invariant bilinear form on the expanded algebra can be constructed using the semigroup product law \eqref{slaw}. This yields
\be\label{invT}
\langle \bt^{(i)}_a \,\bt^{(j)}_b\rangle =\mu_{i\diamond j}\delta_{ab}\,,\quad
\langle \bt^{(i)} \, \bt^{(j)}\rangle = \nu_{i\diamond j} \,,\quad 
\langle \bt^{(i)} \, \btau^{(j)}\rangle = \mu_{i\diamond j}\,,
\ee
where $\mu_{i\diamond j}$ and $\nu_{i\diamond j}$ are two sets arbitrary constants.

\subsection{Expanded Chern-Simons form}
\label{cs}

The expansion method can be used to define an expanded Chern-Simons form. This can be done by defining a connection on $S\times\nw$,
\be
\bA=\theta^a_{(i)}\, \bt_a^{(i)} + \alpha_{(i)}\, \bt^{(i)} + \beta_{(i)}\,\btau^{(i)}\,,\label{connection}
\ee
and using the invariant tensor \eqref{invT}. Equation \eqref{csform} then leads to
\begin{equation}\label{CS_Generic_Lagrangian}
\mathcal{L}_{ S\times\nw}=\mu_{i\diamond j}\left(\delta_{ab}\theta^a_{(i)} \wedge R^b_{(j)}+\alpha_{(i)}\wedge\ed\beta_{(j)}+\beta_{(i)} \wedge \ed\alpha_{(j)}\right)+ \nu_{i\diamond j}\alpha_{(i)} \wedge \ed\alpha_{(j)}\,,
\end{equation}
where we have defined the curvatures associated to the $\theta^a_{(i)}$ gauge fields as
\be
R_{(i)}^a = \ed\theta_{(i)}^a+\epsilon^a_{\;\;b}\, \delta_i^{j\diamond k}\theta_{(k)}^b\, \wedge \alpha_{(j)} \,.\label{Scurvatures}
\ee
In the following, we will consider particular semigroups to Expand the Nappi-Witten symmetry, which will lead to different NR symmetries in 2+1 dimensions.

\section{Generalized Bargmann symmetries}
\label{genbarg}

Let us consider the following expansion of the Nappi-Witten algebra 
\be\label{genbargmann}
S_E^{(n)}\times\nw \,,
\ee
where the semigroup $S_E^{(n)}=\left\{\lambda_0,\lambda_1,\dots,\lambda_{n+1} \right\}$ has a multiplication rule \eqref{slaw} given by
\be
i\diamond j =\left\{
\begin{array}{lll}\label{i+j}
i +j  \quad & \mathrm{if}\quad i+j \leq n\,,
&  \\
n+1 \quad & \mathrm{if}\quad i+j >n\,. &
\end{array}%
\right.
\ee
The element $\lambda_{n+1}$ is the \emph{zero} of the semigroup in the sense that $\lambda_i\cdot\lambda_{n+1}=\lambda_{n+1}$ for any $\lambda_i\in S_E^{(n)}$. The construction of the expanded algebra in this case follows directly from equation \eqref{SexpCom}. It is important to remark that, as explained in \cite{Izaurieta:2006zz}, the fact that the semigroup $S_E^{(n)}$ has a zero element requires to implement the so-called reduction procedure, which sets
\be
\bt^{(n+1)}=\bt_a^{(n+1)}=\btau^{(n+1)}\equiv 0\,,
\ee
together with the condition
\be
\mu_{n+1}\equiv 0
\ee
in the expression \eqref{invT} for the invariant tensor of the expanded algebra.

As we will see in the following, these expanded symmetries are of Bargmann type. Therefore, we will refer to them as \emph{Generalized Bargmann algebras}.

\subsection{Extended Bargmann algebra}
\label{extbarg}
Now we will show that extended Bargmann algebra in 2+1 dimensions is the expanded $\nw$ algebra \eqref{genbargmann} for $n=1$. In order to see this, let us denote the expanded generators \eqref{expandedgen} by
\begin{equation}\label{RenameGen}
\begin{tabular}{lll}
$\bG_a=\bt_a^{(0)} \,\,,$ & $ \bJ=\bt^{(0)}\,,$ & $\text{%
\ } \bS=-\btau^{(0)}\,,$ \\
$\bP_a=\bt_b^{(1)}\,,$ & $ \bH=\bt^{(1)}\,,$ & $\bM=-\btau^{(1)}\,.$
\end{tabular}%
\end{equation}
The expanded algebra \eqref{SexpCom} then takes the form
\be \label{ExtBargAlg}
\begin{array}{lll}
    \left[\bJ,\bG_a\right]=\epsilon_{a}^{\;\;b}\bG_b\,,\quad&
    \left[\bJ,\bP_a\right]=\epsilon_{a}^{\;\;b}\bP_b\,,\quad& \left[\bH,\bG_a\right]=\epsilon_{a}^{\;\;b}\bP_b\,,\\[5pt]
   \left[\bG_a,\bG_b\right]=-\epsilon_{ab}\bS\,,\quad& \left[\bG_a,\bP_b\right]=-\epsilon_{ab}\bM\,,\quad&
\end{array}
\ee
which is the double central extension of the Galilei algebra in 2+1 one dimensions \cite{Bergshoeff:2016lwr,Hartong:2016yrf}, as stated.

Using equation \eqref{invT} we get the corresponding invariant tensor:
\begin{equation}\label{invbarg}
\begin{array}{llll}
\langle \bG_a \bG_b\rangle=\mu_0\delta_{ab}\,, \quad& \langle \bJ \bJ\rangle=\nu_0\,,\quad& \langle \bJ \bS\rangle=-\mu_0\,,\quad&\langle \bH \bS \rangle=-\mu_1\,, \\[6pt]
\langle \bG_a \bP_b\rangle=\mu_1\delta_{ab} \,,\quad& \langle \bJ \bH\rangle=\nu_1\,,\quad&\ \langle \bJ \bM \rangle=-\mu_1\,,\quad& 
\end{array}%
\end{equation}
while the Lagrangian for Extended Bargmann gravity in 2+1 dimensions follows from the Chern-Simons form \eqref{CS_Generic_Lagrangian}. In fact, relabelling the expanded gauge fields \eqref{connection} in the form
\be\label{changeFields}
\begin{array}{lll}
\theta_{(0)}^a= \omega^a\,,&\quad \alpha_{(0)}= \omega\,,&\quad
\beta_{(0)} =-s\,,\\[5pt]
\theta_{(1)}^a= e^a\,,&\quad \alpha_{(1)}= h\,,&\quad
\beta_{(1)}= -m \,,
\end{array}
\ee 
the expanded Chern-Simons Lagrangian takes the form\footnote{Equation \eqref{barggrav} matches the Lagrangian found in \cite{Hartong:2016yrf} after some integration by parts and for $c_5=0$.}\cite{Papageorgiou:2009zc,Hartong:2016yrf}
\be\label{barggrav}
\begin{array}{lll}
    \mathcal{L}_{\rm EB}&=&\mu_0\left[\omega_a\wedge R_{(0)}^a-\omega\wedge\ed s-s\wedge\ed\omega\right]+\nu_0\,\omega\wedge\ed\omega  +\nu_1\Big[\omega\wedge\ed h+h\wedge\ed\omega\Big]\\[5pt]
  &&+\mu_1\left[e_a\wedge R_{(0)}^a+\omega_a\wedge R_{(1)}^a-h\wedge\ed s-s\wedge\ed h-\omega\wedge\ed m-m\wedge\ed\omega\right]\,,
\end{array}
\ee
where the curvatures are given by
\be\label{cbargmann}
\begin{array}{lll}
R_{(0)}^a&=&\ed\omega^a+\epsilon^a_{\;\; b}\omega^a\wedge\omega\,,\\[5pt]
R_{(1)}^a&=&\ed e^a+\epsilon^a_{\;\; b}\left(e^b\wedge\omega+\omega^b\wedge h\right)\,.
\end{array}
\ee

\subsection{Maxwellian Exotic Bargmann algebra}
\label{nonrelmax}
Let us consider now the expanded algebra \eqref{genbargmann} for $n=2$. In this case, we supplement equations \eqref{RenameGen} and \eqref{changeFields} with
\begin{equation}\label{RenameGenmaxwell}
\bZ_a=\bt_b^{(2)}\,,\quad \bZ=\bt^{(2)}\,,\quad \bT=-\btau^{(2)}\,,
\end{equation}
and
\begin{equation}\label{RenameGenmaxwell2}
\theta_{(2)}^a = k^a\,,\quad\alpha_{(2)} = k\,,\quad\beta_{(2)} =-t\,.
\end{equation}
The semigroup product rule \eqref{i+j} then leads to an expanded algebra, whose commutation relations are given by \eqref{ExtBargAlg} plus
\begin{equation}\label{ExtMaxBargAlg}
\begin{array}{lll}
    \left[\bJ,\bZ_a\right]=\epsilon_a^{\;\;b}\bZ_b\,,\quad&
    \left[\bH,\bP_a\right]=\epsilon_a^{\;\;b}\bZ_b\,,\quad&
    \left[\bZ,\bG_a\right]=\epsilon_a^{\;\;b}\bZ_a\,,\\[5pt]
    \left[\bP_a,\bP_b\right]=-\epsilon_{ab}\bT\,,\quad&
    \left[\bG_a,\bZ_b\right]=-\epsilon_{ab}\bT\,.\quad&
\end{array}
\end{equation}
This corresponds to the \emph{Maxwellian Exotic Bargmann} algebra introduced in \cite{Aviles:2018jzw} as a novel NR limit of the Maxwell symmetry in 2+1 dimensions. The invariant tensor can be directly obtained by means of \eqref{invT}, and yields equation \eqref{invbarg} plus
\be\label{invmax}
\begin{array}{llll}
  \langle\bG_a\bZ_b\rangle=\mu_2\delta_{ab} \,,\quad&
  \langle\bJ\bZ\rangle=\nu_2 \,,\quad &
  \langle\bJ\bT\rangle=-\mu_2\,, \quad &
  \langle\bZ\bS\rangle=-\mu_2 \,,\\[5pt]

\langle\bP_a\bP_b\rangle=\mu_2\delta_{ab}\,,\quad&
\langle\bH\bH\rangle=\nu_2 \,, \quad &
\langle\bH\bM\rangle=-\mu_2 \,.\quad &
\end{array}
\ee
In this case, the expanded Chern-Simons form corresponds to the Lagrangian for Maxwellian Exotic Bargmann gravity \footnote{In reference \cite{Aviles:2018jzw}, the Chern-Simons action for Maxwellian Exotic Bargmann gravity is constructed considering \eqref{invbarg} and \eqref{invmax} with $\nu_i=0$. } \cite{Aviles:2018jzw}
\be
\begin{array}{lll}\label{meblag}
 \mathcal{L}_{\rm MEB}
&=&\mathcal{L}_{\rm EB}+\mu_2 \Big[ \omega_a\wedge R_{(2)}^a+e_a\wedge R_{(1)}^a+k_a\wedge R_{(0)}^a \\
&&-\omega\wedge\ed t-t\wedge\ed\omega-h\wedge\ed m-m\wedge\ed h-k\wedge\ed s-s\wedge\ed k\Big]\\
&&+\nu_2\Big[\omega\wedge\ed k+k\wedge\ed\omega+h\wedge\ed h\Big]\,,
\end{array}
\ee
where curvatures $R_{(0)}^a$ and $R_{(1)}^a$ are given in \eqref{cbargmann}, while $R_{(2)}^a$ has the form
\be\label{cmaxwellian}
    R_{(2)}^a=\ed k^a+\epsilon^a_{\;\;b}\left(k^b\wedge\omega+e^b\wedge h+\omega^b\wedge k\right).
\ee

\subsection{Non-relativistic Hietarinta algebra in 2+1 dimensions}
\label{nonrelhiet}
The expansion of the Nappi-Witten algebra by $S_{E}^{(2)}$ can lead to another interesting NR symmetry in 2+1 dimensions. This is given by the following choice of expanded generators
\begin{equation}\label{renaminghietarinta}
\begin{array}{lll}
\bG_a=\bt^{(0)}_{a}\,,\quad &  \bJ=\bt^{(0)}\,,\quad & \bS=-\btau^{(0)}\,, \\
\bZ_a=\bt^{(1)}_{a}\,,\quad &  \bZ=\bt^{(1)}\,,\quad & \bT=-\btau^{(1)} \,, \\
\bP_a=\bt^{(2)}_{a}\,,\quad &  \bH=\bt^{(2)}\,,\quad & \bM=-\btau^{(2)}\,.
\end{array}
\end{equation}
The expanded Nappi-Witten algebra \eqref{SexpCom} now takes the form
\begin{equation}\label{hietalg}
\begin{array}{lll}
    \left[\bG_a,\bG_b\right]=\epsilon_{ab}\bS\,,&\quad&
    \left[\bG_a,\bZ_b\right]=\epsilon_{ab}\bT\,,\\[5pt]
    \left[\bZ_a,\bZ_b\right]=\epsilon_{ab}\bM\,,&\quad&
    \left[\bG_a,\bP_b\right]=\epsilon_{ab}\bM\,,\\[5pt]
    \left[\bJ,\bG_a\right]=-\epsilon_a^{\;\;b}\bG_a,&\quad&
    \left[\bJ,\bZ_a\right]=-\epsilon_a^{\;\; b}\bZ_b\,,\\[5pt]
    \left[\bJ,\bP_a\right]=-\epsilon_a^{\;\; b}\bP_b\,,&\quad&
    \left[\bZ,\bG_a\right]=-\epsilon_a^{\;\; b}\bZ_b\,,\\[5pt]
    \left[\bZ,\bZ_a\right]=-\epsilon_a^{\;\; b}\bP_b\,,&\quad&
    \left[\bH,\bG_a\right]=-\epsilon_a^{\;\; b}\bP_a\,,
\end{array}
\end{equation}
which corresponds to a NR version of the bosonic Hietarinta algebra in 2+1 dimensions \cite{Hietarinta:1975fu}. Unlike the Maxwell case, identifying translations in the way \eqref{renaminghietarinta} leads to Abelian translations. Therefore, even though this algebra can be brought to the Maxwellian exotic Bargmann algebra given in \eqref{ExtBargAlg} and \eqref{ExtMaxBargAlg} by a redefinition of the generators, they are physically different. The Chern-Simons Lagrangian invariant under \eqref{hietalg} defines a non-relativistic limit of the Hietarinta Chern-Simons gravity studied in \cite{Bansal:2018qyz,Chernyavsky:2019hyp}, pretty much in the same way as the Chern-Simons form \eqref{meblag} describes a NR limit of Maxwell gravity\footnote{The analogy between the Hietarinta algebra and the Maxwell algebra relies on the fact that both share the same extended semi-direct product structure \cite{Salgado-Rebolledo:2019kft}.} \cite{Aviles:2018jzw}.
The expanded invariant tensor \eqref{invT} can be read off from \eqref{invmax} by interchanging $\bP_a \leftrightarrow \bZ_a$, $\bH \leftrightarrow \bZ$, $\bM \leftrightarrow \bT$, and the Chern-Simons form \eqref{CS_Generic_Lagrangian} is constructed using
\be\label{changeFields}
\begin{array}{lll}
\theta_{(0)}^a= \omega^a\,,&\quad \alpha_{(0)}= \omega\,,&\quad
\beta_{(0)} =-s\,,\\[5pt]
\theta_{(1)}^a= k^a\,,&\quad \alpha_{(1)}= k\,,&\quad
\beta_{(1)}= -t \,,\\[5pt]
\theta_{(2)}^a = e^a\,,&\quad\alpha_{(2)} = h\,,&\quad\beta_{(2)} =-m\,.
\end{array}
\ee 
The final result is
\be\label{lagrangianhiet}
\begin{array}{lll}
    \mathcal{L}_{\rm NRH} &=&
    \mu_0\left[\omega_a\wedge R_{(0)}^a-\omega\wedge\ed s-s\wedge\ed\omega\right]+
    \nu_0\;\omega\wedge\ed\omega\\[5pt]
    &&+\mu_1\left[k_a\wedge R_{(0)}^a+\omega_a\wedge R_{(1)}^a-k\wedge\ed s-s\wedge\ed k-\omega\wedge\ed t-t\wedge\ed\omega\right]\\[5pt]
    &&+\nu_1\Big[\omega\wedge\ed k+k\wedge\ed\omega\Big]+\mu_2\Big[\omega_a\wedge R_{(2)}^a+k_a\wedge R_{(1)}^a+e_a\wedge R_{(0)}^a \\[5pt]
    &&-\omega\wedge\ed m-m\wedge\ed\omega-k\wedge\ed t-
    t\wedge\ed k-h\wedge\ed s-s\wedge\ed h\Big]\\[5pt]
    &&+\nu_2\Big[\omega\wedge\ed h+h\wedge\ed\omega+k\wedge\ed k\Big],
\end{array}
\ee
where
\be
\begin{array}{lll}
    R_{(0)}^a&=&\ed\omega^a+\epsilon^a_{\;\; b}\omega^b\wedge\omega\,,\\[5pt]
    R_{(1)}^a&=&\ed k^a+\epsilon^a_{\;\; b}\left(k^b\wedge\omega+\omega^b\wedge k\right)\,,\\[5pt]
    R_{(2)}^a&=&\ed e^a+\epsilon^a_{\;\; b}\left(e^b\wedge\omega+\omega^b\wedge h+k^b\wedge k\right)\,.
\end{array}
\ee
This theory can be obtained by means of an In\"{o}n\"{u}-Wigner contraction of the relativistic Hietarinta symmetry studied in \cite{Bansal:2018qyz,Chernyavsky:2019hyp} by incorporating three $\mathfrak{u}(1)$ gauge fields, in the same way as it happens in the Maxwell case \cite{Aviles:2018jzw}.

\section{Generalized Newton-Hooke symmetries}
\label{gennh}
In this section we define a different kind of expanded Nappi-Witten algebras, which we will refer to as \emph{Generalized Newton-Hooke  algebras}. They are are given expansions of the form
\be\label{gennewtonhooke}
S_M^{(n)}\times\nw \,,
\ee
where $S_M^{(n)}={\lambda_0 ,\lambda_1,\dots,\lambda_n}$ denotes a semigroup with product \eqref{slaw} given by
\begin{equation} \label{smn}
i\diamond j=\left\{
\begin{array}{lll}
\alpha +\beta \,\,\,\, & \mathrm{if}\,\,\,\,\alpha +\beta \leq n\,,
&  \\
\alpha +\beta -2\left[ \frac{n+1}{2}\right]\,\,\, & \mathrm{if}%
\,\,\,\,\alpha +\beta >n \,.&
\end{array}%
\right. 
\end{equation}
This type of algebras can be related to the Generalized Bargmann algebras studied in Section \ref{genbarg} by means of an In\"on\"u-Wigner contraction \cite{inonu1993contraction}. This can be made by introducing a parameter $\ell$ and rescaling the Lie algebra generators, the gauge fields, and the invariant tensor constants in the form
\be\label{contraction}
\begin{array}{lll}
\bt_a^{(i)}\rightarrow\ell^i \bt_a^{(i)}\,,&\quad 
\bt^{(i)}\rightarrow\ell^i \bt^{(i)}\,,&\quad 
\btau^{(i)}\rightarrow\ell^i \btau^{(i)}\,,\\[5pt]

\theta^a_{(i)}\rightarrow\dfrac{1}{\ell^i}\theta^a_{(i)}\,,&\quad 
\alpha_{(i)}\rightarrow\dfrac{1}{\ell^i} \alpha_{(i)}\,,&\quad 
\beta_{(i)}\rightarrow\dfrac{1}{\ell^i} \beta_{(i)}\,,\\[10pt]

\mu_i\rightarrow\ell^i \mu_i\,,&\quad 
\nu_i\rightarrow\ell^i \nu_i\,.&
\end{array}
\ee
In the limit $\ell\rightarrow\infty$, the expanded algebra \eqref{gennewtonhooke} contracts to \eqref{genbargmann}, i.e.
\begin{equation}
S_M^{(n)}\times \nw \xrightarrow[\ell\rightarrow\infty]\, S_E^{(n)}\times \,.\nw
\end{equation}
As we will see in the following examples, the limit can be consistently applied at the level of the corresponding Chern-Simons gravity actions. 

\subsection{Extended Newton-Hooke algebra}
\label{expnh}
Let us consider now the expanded algebra \eqref{gennewtonhooke} for $n=1$, i.e. the direct product of the Nappi-Witten algebra and the semigroup $S_M ^{(1)}\backsimeq\mathbb{Z}_2$. Using the prescription \eqref{contraction} and labelling the expanded generators in the form \eqref{RenameGen}, we find the extended Newton-Hooke algebra \cite{Hartong:2016yrf}, given by the commutation relations \eqref{ExtBargAlg} plus
\begin{equation}\label{generatorsnh}
    \left[\bP_a,\bP_b\right]=-\frac{1}{\ell^2}\epsilon_{ab}\;\bS \,, \quad
    \left[\bH,\bP_a\right]=\frac{1}{\ell^2}\epsilon_{a}^{\;\;b}\bG_b \,.
\end{equation}
Similarly, applying \eqref{contraction} to the definition of the invariant tensor \eqref{invT} yields \eqref{invbarg} plus
\be
 \langle\bP_a\bP_b\rangle=\frac{\mu_0}{\ell^2}\delta_{ab}\,,\quad
 \langle\bH\bH \rangle=\frac{\nu_0}{\ell^2}\,,\quad
 \langle\bH\bM\rangle=-\frac{\mu_0}{\ell^2}  \,.
\ee
We can see that in this case the parameter $\ell$ can be identified with the AdS radius. The corresponding Chern-Simons Lagrangian \eqref{CS_Generic_Lagrangian} can be put in a familiar form by first applying \eqref{contraction} and then \eqref{changeFields}. This gives \cite{Papageorgiou:2010ud,Hartong:2016yrf}
\be
\begin{array}{lll}
\mathcal{L}_{\rm ENH} &= & \mathcal{L}_{\rm EB}+\frac{\mu_0}{\ell^2}\left[  e_a\wedge R_{(1)}^a+\epsilon_{ab}\;\omega^a \wedge e^b \wedge h -h\wedge\ed m -m\wedge\ed h\right] \\ [5pt]
&& +\frac{\mu_1}{\ell^2} \epsilon_{ab}\; e^a \wedge e^b \wedge h +\frac{\nu_0}{\ell^2}h\wedge\ed h\,,
\end{array}
\ee
where $\mathcal{L}_{\rm EB}$ is given by \eqref{barggrav} and we have also used \eqref{cbargmann}.

This Lagrangian reduces to \eqref{barggrav} in the limit $\ell\rightarrow\infty$, as expected. As explained in \cite{Alvarez:2007ys,Hartong:2017bwq,Joung:2018frr}, the Newton-Hooke algebra is isomorphic to a direct product of two copies of the Nappi-Witten algebra. This can be easily seen by defining the following combination of generators
\be\label{2nw}
\bt^{\pm}_a=\frac{1}{2}\left( \bG_a \pm  \ell \bP_a \right)\,,\quad
\bt^{\pm}=\frac{1}{2}\left( \bJ \pm  \ell \bH \right)\,,\quad
\btau^{\pm}=-\frac{1}{2}\left( \bS \pm  \ell \bM \right)\,,
\ee
which satisfy the commutation relations \eqref{nwalg}.

\subsection{Newton-Hooke $\oplus\,\nw$}
\label{nonreladsl}

In the $n=2$ case, the expanded Nappi-Witten algebra  \eqref{gennewtonhooke} corresponds to a NR version of the semi-simple extension of the Poincar\'e symmetry \cite{Soroka:2006aj,Soroka:2011tc} in 2+1 dimensions. This extension is usually referred to as AdS-Lorentz algebra, for it defines a deformation of the Maxwell symmetry isomorphic to $\mathfrak{so}(2,2)\oplus\mathfrak{so}(2,1)$ \cite{Gomis:2009dm}. As in the previous case, we first implement \eqref{contraction} and subsequently label the expanded generators in the form \eqref{RenameGen} and \eqref{RenameGenmaxwell}. This leads the commutation relations \eqref{ExtBargAlg}, \eqref{ExtMaxBargAlg} and
\begin{equation}\label{nonreladslalg}
\begin{array}{lll}
    \left[\bZ,\bZ_a\right]=\frac{1}{\ell^2}\epsilon_{a}^{\;\;b} \bZ_b\,,\quad&
    \left[\bP_a,\bZ_b\right]=-\frac{1}{\ell^2}\epsilon_{ab} \bM \,, \quad&
    \left[\bZ_a,\bZ_b\right]=-\frac{1}{\ell^2}\epsilon_{ab}\bT \,,\\[5pt]
   \left[\bH,\bZ_a\right]=\frac{1}{\ell^2}\epsilon_{a}^{\;\;b} \bP_b\,,\quad&
    \left[\bT,\bP_a\right]=\frac{1}{\ell^2}\epsilon_{a}^{\;\;b}\bP_b  \,.\quad&
\end{array}
\end{equation}
This novel non-relativistic algebra reduces to the Maxwellian Exotic Bargmann symmetry \cite{Aviles:2018jzw} for $\ell\rightarrow\infty$. In the same way as the AdS-Lorentz symmetry in 2+1 dimensions is isomorphic to $\mathfrak{so}(2,2)\oplus\mathfrak{so}(2,1)$, the algebra \eqref{nonreladslalg} is isomorphic to the semi-direct sum of the Extended Newton-Hooke algebra and the Nappi-Witten algebra. In order to see this, one can apply the same kind of redefinition for the generators that is used in the relativistic case (see, for instance, \cite{Concha:2018jjj}) and define
\be\label{enh2}
\begin{array}{lll}
\bar{\bG}_a= \ell^2\bZ_a\,,\quad& \bar{\bJ}= \ell^2\bZ \,,&\quad \bar{\bS}= \ell^2\bT \,,\\[6pt]
\bar{\bP}_a= \bP_a\,,\quad& \bar{\bH}= \bH \,,&\quad \bar{\bM}= \bM \,.
\end{array}
\ee
These generators satisfy the Extended Newton-Hooke algebra \eqref{generatorsnh}, while the combinations
\be
\bar{\bt}_a= \bG_a-\ell^2\bZ_a \,,\quad  \bar{\bt}= \bJ -\ell^2\bZ_a  \,,\quad   \bar{\btau}=\bS-\ell^2\bZ_a  \,,
\ee
commute with \eqref{enh2} and form a Nappi-Witten subalgebra with commutation relations \eqref{nwalg}. Furthermore, if one applies the change of basis \eqref{2nw} to the generators \eqref{enh2}, one can express \eqref{nonreladslalg} as three copies of the Nappi-Witten algebra. This is in complete analogy with the fact that, in 2+1 dimensions, the AdS Lorentz algebra can be written as three copies of the $\mathfrak{so}(2,1)$ algebra \cite{Concha:2018jjj}.

The invariant tensor \eqref{invT} in this case takes the form
\be
\begin{array}{llll}

\langle\bP_a\bZ_b\rangle=\frac{\mu_1}{\ell^2}\delta_{ab} \,,\quad&
\langle\bH\bZ \rangle=\frac{\nu_1 }{\ell^2} \,,\quad&
\langle\bH\bT\rangle=-\frac{\mu_1}{\ell^2} \,,\quad&  \langle\bZ\bM\rangle=-\frac{\mu_1}{\ell^2}\,,\\[5pt]

\langle\bZ_a\bZ_b\rangle=\frac{\mu_2}{\ell^2}\delta_{ab} \,,\quad&
\langle\bZ\bZ \rangle=\frac{\nu_2}{\ell^2} \,,\quad &
 \langle\bZ\bT \rangle=-\frac{\mu_2}{\ell^2} \,,\\[5pt]

\end{array}
\ee
and the expanded Chern-Simons form \eqref{CS_Generic_Lagrangian} can be written as a gravity theory by applying the prescription \eqref{contraction} followed by \eqref{changeFields} and \eqref{RenameGenmaxwell2}:
\be
\begin{array}{lll}
    \mathcal{L}_{{\rm ENH}\oplus\nw}&=& \mathcal{L}_{\rm MEB}+

\frac{\mu_1}{\ell^2}\Big[ e_a\wedge R_{(2)}^a +k_a\wedge R_{(1)}^a+ \epsilon_{ab}\left(\omega^a  +\frac{1}{\ell^2} k^a \right)
\wedge \left(e^b\wedge k+k^b\wedge h\right)  \\[5pt]

&&+\frac{1}{\ell^2} \epsilon_{ab} \,e^a\wedge k^b\wedge k -k\wedge\ed m-m\wedge\ed k-h\wedge\ed t-t\wedge\ed h \Big]  \\[5pt]

&&+
\frac{\mu_2}{\ell^2}\Big[ \epsilon_{ab}
 \left(\omega^a  +\frac{1}{\ell^2} k^a \right)
\wedge k^b\wedge k  + \epsilon_{ab}\, e^a\wedge \left(e^b\wedge k+k^b\wedge h \right)+ k_a\wedge R_{(2)}^a \\[5pt]
&& - k\wedge\ed t- t\wedge\ed k \Big] 
+\frac{\nu_1}{\ell^2}\left( k\wedge\ed h+ h\wedge\ed k\right)+\frac{\nu_2}{\ell^2}\;k\wedge\ed k \,,\\[5pt]

\end{array}
\ee
where $ \mathcal{L}_{{\rm MEB}}$ is given by \eqref{meblag} and for simplicity we have expressed the  Lagrangian in terms of the curvatures \eqref{cbargmann} and \eqref{cmaxwellian} obtained in the Maxwellian Exotic Bargmann case. Naturally, in the limit $\ell\longrightarrow\infty$ this Lagrangian reduces to the one for Maxwellian Exotic Bargmann gravity \eqref{meblag}.

\section{Carroll algebra}
\label{carroll}

The Carroll symmetry corresponds to the ultrarelativistic limit of the Poincar\'e algebra \cite{cartan1923varietes,cartan1924varietes} (for a geometric description see\cite{Duval:2014uoa}). In this section, we show how to obtain the Carroll algebra as an expanded Nappi-Witten algebra. In order to do this we will consider a rather unusual semigroup $\tilde{S}^{(3)}$, whose multiplication rule is given by
\begin{equation}
\begin{tabular}{l|lllll}
$\cdot$ & $\lambda _{0}$ & $\lambda _{1}$ & $\lambda _{2}$ & $\lambda _{3}$ & $\lambda _{4}$\\ \hline
$\lambda _{0}$ & $\lambda _{0}$ & $\lambda _{1}$ & $\lambda _{2}$ & $\lambda _{3}$ & $\lambda _{4}$ \\
$\lambda _{1}$ & $\lambda _{1}$ & $\lambda _{4}$ & $\lambda _{3}$ & $\lambda _{4}$ & $\lambda _{4}$ \\
$\lambda _{2}$ & $\lambda _{2}$ & $\lambda _{3}$ & $\lambda _{4}$ & $\lambda _{4}$ & $\lambda _{4}$ \\
$\lambda _{3}$ & $\lambda _{3}$ & $\lambda _{4}$ & $\lambda _{4}$ & $\lambda _{4}$ & $\lambda _{4}$ \\
$\lambda _{4}$ & $\lambda _{4}$ & $\lambda _{4}$ & $\lambda _{4}$ & $\lambda _{4}$ & $\lambda _{4}$ \,,
\end{tabular}
\label{sml}
\end{equation}%
This semigroup differs from the standard $S_{E}^{(3)}$ in that now $\lambda_1\cdot\lambda_1=\lambda_4$, where $\lambda_4$ is the zero of the semigroup. Also, in this case, the expansion mechanism requires to apply the resonant reduction prescription introduced in \cite{Izaurieta:2006zz} with the following subspace decomposition of the Nappi-Witten algebra
\be
V_0=\left\{\ \bt \right\} \,,\quad V_1=\left\{\ \bt_a \right\} \,,\quad V_2=\left\{\ \btau \right\} \,,
\ee
which can be shown to be resonant with the following decomposition of $\tilde{S}^{(3)}$:
\be\label{carrollaction}
S_0=\left\{ \lambda_0 , \lambda_4 \right\} \,,\quad S_1=\left\{\lambda_1 , \lambda_2 , \lambda_4 \right\} \,,\quad S_2=\left\{ \lambda_3 , \lambda_4 \right\} \,.
\ee
The generators of the algebra
$
\tilde{S}^{(3)}\times \nw
$
will be denoted by
\begin{align}
    &\bJ=\bt^{(0)}\,,& & \bG_a=\bt_{a}^{(1)}\,\,,& &\bP_a=\bt_{a}^{(2)}\,, & &\bH=-\btau^{(3)}\,, &
\end{align}
which automatically leads to the Carroll algebra \cite{levy1965nouvelle,bacry1968possible}:
\be
   \left[\bJ,\bG_a\right]=\epsilon_{a}^{\;\;b}\bG_b\,,\quad \left[\bJ,\bP_a\right]=\epsilon_{a}^{\;\; b}\bP_b\,, \quad   \left[\bG_a,\bP_b\right]=-\epsilon_{ab}\bH\,.
\ee
Using \eqref{invT} and \eqref{CS_Generic_Lagrangian}, the invariant tensor is found to be
\be
\langle \bG_a \bP_b\rangle=\mu \delta_{ab}\,, \quad \langle \bJ \bJ\rangle=\nu \,,\quad \langle \bJ \bH\rangle=-\mu \,,
\ee
while the corresponding Chern-Simons Lagrangian defines Carroll gravity in 2+1 dimensions:
\begin{equation}\label{invcarroll}
\mathcal{L}_{\rm Carroll}=\mu \left[e_a\wedge R^{a}_{(1)}+\omega_a\wedge R^{a}_{(2)}-\omega\wedge\ed h-h\wedge\ed\omega\right]+\nu\, \omega\wedge\ed\omega \,, 
\end{equation}
where
\be
R_{(1)}^a=\ed\omega^a+\epsilon^a_{\;\;b}\,\omega^b\wedge\omega\,,\quad
R_{(2)}^a=\ed e^a+\epsilon^a_{\;\;b}\,e^b\wedge\omega \,.
\ee
Carollian gravity was first studied in \cite{Hartong:2015xda} in the metric formulation, while \cite{Bergshoeff:2016soe} considered the Chern-Simons formulation for $\nu=0$. The action \eqref{carrollaction} can be obtained as a contraction of the (A)dS-Carrol symmetry \cite{Ravera:2019ize,Matulich:2019cdo}.

\section{Pseudo-Newton-Hooke algebra}
\label{pseudo}
The Nappi-Witten algebra can be defined as particular case of a double extended algebra. The way to see this is the following:
given two Abelian Lie algebras $\mathfrak{g}_1=\sp\{\bX_a \}$, and $\mathfrak{g}_2=\sp\{\bY_\alpha \}$, the double extended algebra $D(\mathfrak{g}_1 , \mathfrak{g}_2)$ is \cite{Figueroa-OFarrill:1994liu}
\be\label{dealg}
\left[ \bX_a , \bX_b \right]= f_{\alpha a}^{\;\;\;b} \bY^{*\alpha}\,,\quad \left[ \bY_\alpha , \bX_a \right]= f_{\alpha a}^{\;\;\;b} \bX_b\,,
\ee
where $\bY^{*\alpha}$ denotes the basis of the dual Lie algebra $\mathfrak{g}_2^*$ with respect to the pairing
\be\label{pairingy}
\left\langle \bY_\alpha\, \bY^{*\beta}\right\rangle=\delta^\beta_\alpha \,.
\ee
The constants $f_{\alpha i}^{\;\;\;j}$ that define the action of $\mathfrak{g}_2$ on $\mathfrak{g}_1$ must satisfy the anti-symmetry condition
\be\label{symmcond}
f_{\alpha a}^{\;\;\;b}g_{bc} =
-f_{\alpha c}^{\;\;\;b}g_{ab} \,,
\ee
where $g_{ab}$ is a non-degenerate metric on $\mathfrak{g}_1$
\be\label{metricx}
\left\langle \bX_a \, \bX_b\right\rangle= g_{ab} \,.
\ee
Then, invariant non-degenerate bilinear form on $D(\mathfrak{g}_1,\mathfrak{g}_2)$ can be constructed out of \eqref{pairingy} and \eqref{metricx}, together with a non-degenerate metric $g_{\alpha \beta}=\left\langle \bY_\alpha \, \bY_\beta \right\rangle$ on $\mathfrak{g}_2$.

Based on the previous discussion, it is straightforward to see that the Nappi-Witten algebra \eqref{nwalg} is the Double extended algebra \be\nw=D\left(\mathbb{R}^2,\mathfrak{u}(1)\right)\,,\ee
where $\mathbb{R}^2$ is the two-dimensional translation algebra endowed with the Euclidean metric
\be
\bX_a=\bt_a=\left\{\bt_1 , \bt_2 \right\}\,,\quad g_{ab}=\delta_{ab} \,.
\ee
Denoting the $\mathfrak{u}(1)$ generator and its dual by
\be
\bY_1=\bt \,,\quad \bY^{*1}=\btau \,,\quad \left\langle \bY_1 \, \bY^{*1}\right\rangle=1 \,,\quad g_{11}=\nu \,,
\ee
the invariant tensor \eqref{nwmetric} is recovered and \eqref{dealg} leads to the commutation relations \eqref{nwalg}. The action of $\mathfrak{u}(1)$ on $\mathbb{R}^2$ is determined by $f_{1a}^{\;\;\;b}=\epsilon_a^{\;b}$, which clearly satisfies \eqref{symmcond}.

The interesting point here is that one can alternatively consider $\mathfrak{g}_1=R^{1,1}$ as the two-dimensional translation algebra endowed with the Minkowski metric
\be 
g_{ab}=\eta_{ab}={\rm diag}(-+)\,,
\ee
and define the \emph{pseudo-Nappi-Witten algebra} 
\be\label{pseudonwalg}
\pnw=D\left(R^{1,1},\mathfrak{u}(1)\right) \,.
\ee
In this case, the commutation relations \eqref{dealg} take the same form as \eqref{nwalg}, and the conditions \eqref{symmcond} are automatically fulfilled provided indices are now lowered with the two-dimensional Minkowski metric. The invariant form \eqref{nwmetric}, however,  has to be replaced by
\be\label{pnwmetric}
    \langle \bt_a\, \bt_b\rangle = \eta_{ab}\,,\quad
    \langle \bt\,\btau\rangle = 1 \,,\quad
    \langle \bt\,\bt\rangle = \nu\, . 
\ee
The algebra \eqref{pseudonwalg} defines a central extension of the Poincar\'e algebra in 1+1 dimensions \cite{Cangemi:1992up,Cangemi:1992bj}.

As we did in Section \ref{exph}, we can define expansions of the $\pnw$ algebra by considering its direct product with a semigroup of the form \eqref{slaw},
\be
S\times\pnw = \sp\big\{ \left\{\lambda_i\right\}\otimes  \left\{\bt,\bt_a,\btau  \right\}\big\} \,.
\ee
The generators of the expanded Lie algebra will satisfy commutation relations similar to \eqref{SexpCom}, but endowed with the invariant bilinear form:
\be\label{invT2}
    \langle \bt^{(i)}_a \,\bt^{(j)}_b\rangle =\mu_{i\diamond j}\eta_{ab} \,,\quad
    \langle \bt^{(i)} \, \btau^{(j)}\rangle = \mu_{i\diamond j} \,,\quad
    \langle \bt^{(i)} \, \bt^{(j)}\rangle = \nu_{i\diamond j}\,.
\ee

As an example, we can consider an expansion with the semigroup $S_M^{(1)}$, given in \eqref{smn}. In this case, we will consider the following redefinition for the expanded generators:
\begin{equation}\label{RenameGen3}
\begin{tabular}{lll}
$ \bG_a= \epsilon_{a}^{\;\;b} \bt_b^{(0)} \,\,,$ & $ \bJ=\bt^{(0)}\,,$ & $\text{%
\ } \bS=-\btau^{(0)}\,,$ \\
$\ell\bP_a=\bt_a^{(1)}\,,$ & $ \ell\bH=\bt^{(1)}\,,$ & $\ell\bM=-\btau^{(1)}\,.$
\end{tabular}%
\end{equation}
This leads to the pseudo-Newton-Hooke symmetry \cite{Hartong:2017bwq}
\begin{equation}\label{pseudonh}
\begin{array}{lll}
    \left[\bG_a,\bG_b\right]=\epsilon_{ab}\;\bS,&\quad&
    \left[\bG_a,\bP_b\right]=-\eta_{ab}\;\bM,\\[5pt]
    \left[\bP_a,\bP_b\right]=-\frac{1}{\ell^2}\epsilon_{ab}\;\bS,&\quad&
    \left[\bJ,\bG_a\right]=\epsilon_{a}^{\;\;b}\;\bG_b,\\[5pt]
    \left[\bJ,\bP_a\right]=\epsilon_{a}^{\;\;b}\;\bP_b,&\quad&
    \left[\bH,\bG_a\right]=\bP_a,\\[5pt]
    \left[\bH,\bP_a\right]=\frac{1}{\ell^2}\bG_a\,. &\quad&
\end{array}
\end{equation}
Similarly, an expansion with the semigroup $S_E^{(1)}$ will lead to the pseudo-Bargmann algebra, which also follows from \eqref{pseudonh} in the limit $\ell \rightarrow\infty$. It is also possible to define a pseudo-Maxwellian-Exotic Bargmann algebra, as well as pseudo-Newton-Hooke $\oplus\,\pnw$, by considering the semigroups $S_E^{(2)}$ and $S_M^{(2)}$. In the same way as done with the Nappi-Witten algebra in Sections \ref{genbarg} and \ref{gennh}, this construction can be generalized to obtain two families of expanded algebras, $S_E^{(n)}\times\tilde\nw$ and $S_M^{(n)}\times\tilde\nw$, defining generalized pseudo-Bargmann and pseudo-Newton-Hooke symmetries, respectively.

\section{Conclusion}
\label{conclusions}

In this paper, we have shown how to obtain various NR symmetries in 2+1 dimensions as expansions of the Nappi-Witten algebra. The expansion method based on Abelian semigroups is used to construct the invariant bilinear form and the Chern-Simons action locally invariant under each expanded algebra.

When expanding the Nappi-Witten algebra with the semigroup $S_E^{(n)}$, we obtained a family of NR algebras of Bargmann type. We have worked out explicitly the simplest examples. The $n=1$ case leads to the extended Bargmann algebra in 2+1 dimensions. The invariant form obtained from the Nappi-Witten metric leads to general one obtained in \cite{Hartong:2016yrf} with $c_5=0$, and the corresponding extended Bargmann gravity is reproduced from the expanded Nappi-Witten Chern-Simons action. In the $n=2$ case, we obtained the Maxwellian Exotic Bargmann algebra, which was first constructed in \cite{Aviles:2018jzw} as an In\"{o}n\"u-Wigner contraction of the direct sum of the Maxwell algebra in 2+1 dimensions and three $u(1)$ generators. Furthermore, a different redefinition of the expanded generators leads to an extended non-relativistic version of the Hietarinta algebra in 2+1 dimensions \cite{Chernyavsky:2019hyp}. In both cases, the corresponding gravity theory is obtained from the Nappi-Witten Chern-Simons theory through the expansion method. 

Another family of NR symmetries is obtained when considering the semigroup $S_M^{(n)}$. In this case, the expanded algebras are of Newton-Hooke type. In fact, the simplest case $n=1$ was shown to yield the extended Newton-Hooke symmetry and the corresponding Chern-Simons gravity theory described in \cite{Hartong:2016yrf} (with $c_5=c_2$). For $n=2$ we find a NR version of the semi-simple extension of the Poincar\'e algebra introduced in \cite{Gomis:2009dm,Soroka:2011tc}. Following \cite{Aviles:2018jzw}, this algebra can be understood as a Maxwellian Exotic Newton-Hooke symmetry.

For $n>2$ these two families of expanded Nappi-Witten algebras define central extensions of the Generalized Galilean symmetries presented in \cite{Gonzalez:2016xwo} in the case of three-space time dimensions, and define NR limits of the generalized Poincar\'e and generalized AdS algebras \cite{Concha:2014zsa,Concha:2016kdz}.

Another interesting NR symmetry discussed here is the Carroll algebra. Even though it does not belong to any of the algebra families previously defined, it was obtained as an expansion of the Nappi-Witten using a particular non-conventional semigroup. The bilinear form of the Nappi-Witten algebra can be used to define the Carroll invariant Chern-Simons gravity theory previously considered \cite{Bergshoeff:2016soe,Ravera:2019ize,Matulich:2019cdo}.

Subsequently, we considered a different version of the Nappi-Witten algebra given by the central extension of the Poincar\'e algebra in 1+1 dimensions \cite{Cangemi:1992up,Cangemi:1992bj}. This symmetry is a double extension of the translation algebra endowed with the Minkowski metric. Upon expansion, it leads to two different kinds of relativistic symmetries that generalize the pseudo-Galilei and pseudo-Newton-Hooke algebras introduced in \cite{Hartong:2017bwq} in the context of the AdS/CFT correspondence.

A very appealing future direction is to explore the possibility of obtaining NR super algebras in 2+1 dimensions from a supersymmetric extension of the Nappi-Witten algebra. In this case, the expansion mechanism would allow one to construct the bilinear invariant form and the Chern-Simons action associated to generalized super Bargmann and generalized super Newton-Hooke algebras in a straightforward way, leading to novel NR supergravity theories in 2+1 dimensions.

The Nappi-Witten algebra was used in \cite{Nappi:1993ie} to construct a WZW model based on a non-semi-simple group, leading to plane wave string theory backgrounds (see also \cite{Duval:1994qz}). Along the same lines, it would be interesting to explore WZW models associated with expansions of the Nappi-Witten symmetry here presented, as well as the background metrics that they can provide.

\section*{Acknowledgements}
The authors would like to thank J. Gomis, S. Prohazka and L. Ravera for interesting discussions and useful comments. D.M.P. and P.S-R. acknowledge DI-VRIEA for financial support through Proyecto Postdoctorado 2019 VRIEA-PUCV. 

\bibliographystyle{utphys}

\bibliography{referencesENW}

\end{document}